# Compact Turnkey Soliton Microcombs at Microwave Rates via Wafer-Scale Fabrication

Yuanlei Wang[1,2]*, Ze Wang[1]*, Chenghao Lao[1,2]*, Tianyu Xu[1], Yinke Cheng[1,2], Zhenyu Xie[1], Junqi Wang[1], Haoyang Luo[1], Xin Zhou[2], Bo Ni[1,3], Kaixuan Zhu[1], Yanwu Liu[1], Xing Jin[1], Min Wang[2], Jian-Fei Liu[2], Xuening Cao[2], Ting Wang[2], Qihuang Gong[1,3,4], Bei-Bei Li[2], Fangxing Zhang[3], Yun-Feng Xiao[1,3,4] and Qi-Fan Yang[1,3,4]†

[1]State Key Laboratory for Artificial Microstructure and Mesoscopic Physics and Frontiers Science Center for Nano-optoelectronics, School of Physics, Peking University, Beijing 100871, China

[2]Beijing National Laboratory for Condensed Matter Physics, Institute of Physics, Chinese Academy of Sciences, Beijing 100190, China

[3]Peking University Yangtze Delta Institute of Optoelectronics, Nantong, Jiangsu 226010, China

[4]Collaborative Innovation Center of Extreme Optics, Shanxi University, Taiyuan 030006, China

*These authors contributed equally to this work.

†Corresponding author: leonardoyoung@pku.edu.cn

Soliton microcombs generated in nonlinear microresonators facilitate the photonic integration of timing, frequency synthesis, and astronomical calibration functionalities. For these applications, low-repetition-rate ($f_r$) soliton microcombs are essential as they establish a coherent link between optical and microwave signals. However, the required pump power typically scales with $f_r^{-1}$, and the device footprint scales with $f_r^{-2}$, rendering low-$f_r$ soliton microcombs challenging to integrate within photonic circuits. This study designs and fabricates $Si_3N_4$ microresonators on 4-inch wafers with highly compact form factors. The resonator geometries are engineered from ring to finger and spiral shapes to enhance integration density while attaining quality factors over $10^7$. Driven directly by an integrated laser, soliton microcombs with $f_r$ below 10 GHz are demonstrated via turnkey initiation. The phase noise performance of the synthesized microwave signals reaches -130 dBc Hz$^{-1}$ at 100 kHz offset frequency for 10 GHz carrier frequencies. This work enables the high-density integration of soliton microcombs for chip-based microwave photonics and spectroscopy applications.

## Introduction

Optical frequency combs (OFCs) provide a series of equally spaced spectral lines that bridge optical and microwave frequency domains[1]. Recent advances have reduced OFC systems from bulky, table-top lasers to chip-scale platforms[2,3]. In particular, soliton microcombs generated in high-quality-factor (high-$Q$) microresonators have emerged as a leading solution among integrated comb sources due to their high coherence and broad spectral coverage[4]. Among microresonators based on $MgF_2$[5,6], $SiO_2$[7], $LiNbO_3$[8], $AlN^9$ and $SiC^{10}$, the $Si_3N_4$ platform is particularly compelling, owing to its broad transparency window, high Kerr nonlinearity, and ultra-low optical losses[11–16]. Notably, high-$Q$ $Si_3N_4$ microresonators pumped by laser chips have already demonstrated fully integrated soliton microcombs[17–21]. A crucial next step is to integrate these comb sources with more complex photonic circuits, where a high integration density is essential for reducing costs and achieving advanced functionalities. Such developments can benefit the rapidly growing demands from communicational[22,23] and computational tasks[24].

Different OFC applications demand specific comb spacings, or repetition frequencies ($f_r$), which are typically near the microresonator's free-spectral-range (FSR) defined by the round-trip length ($L$). Soliton microcombs can achieve $f_r$ values ranging from gigahertz[25] to terahertz,[26] which remain challenging for conventional table-top OFCs[1]. Such large comb spacings are compatible with dense-wavelength-division-multiplexing technologies in optical communications[27–29] and are resolvable by spectrometers used in astronomical spectroscopy calibration[30,31] (Figure 1a). However, certain applications—including synthetic aperture radar[32] and chemical absorption spectroscopy[33]—require low repetition frequencies (<10 GHz), demanding resonators with round-trip lengths exceeding centimeters. This creates two key challenges: First, microring resonators with such large $f_r$ can occupy footprints over 5 mm × 5 mm, limiting the density of on-chip comb generators and processing elements; second, the pump power for soliton generation scales with the resonator length, and low-$f_r$ soliton microcombs may consume power exceeding the capabilities of integrated lasers.

Here, we implement compact microresonators with finger-shaped and spiral-shaped geometries[34,35] to generate soliton microcombs. Unlike microrings, whose footprint scales with $L^2$, these alternative designs scale more favorably with $L$, enabling high-density integration of low-$f_r$ soliton microcombs with other functional components, such as processor and sensor (Figure 1b). We fabricate these devices on 4-inch wafers with a 777-nm-thick $Si_3N_4$ waveguide core, achieving intrinsic quality factors ($Q_0$) exceeding $10^7$. By co-integrating a distributed-feedback (DFB) laser with our microresonator, we demonstrate soliton microcombs operating at $f_r < 10$ GHz, featuring low phase noise, straightforward turnkey initiation, and reliable long-term stability.

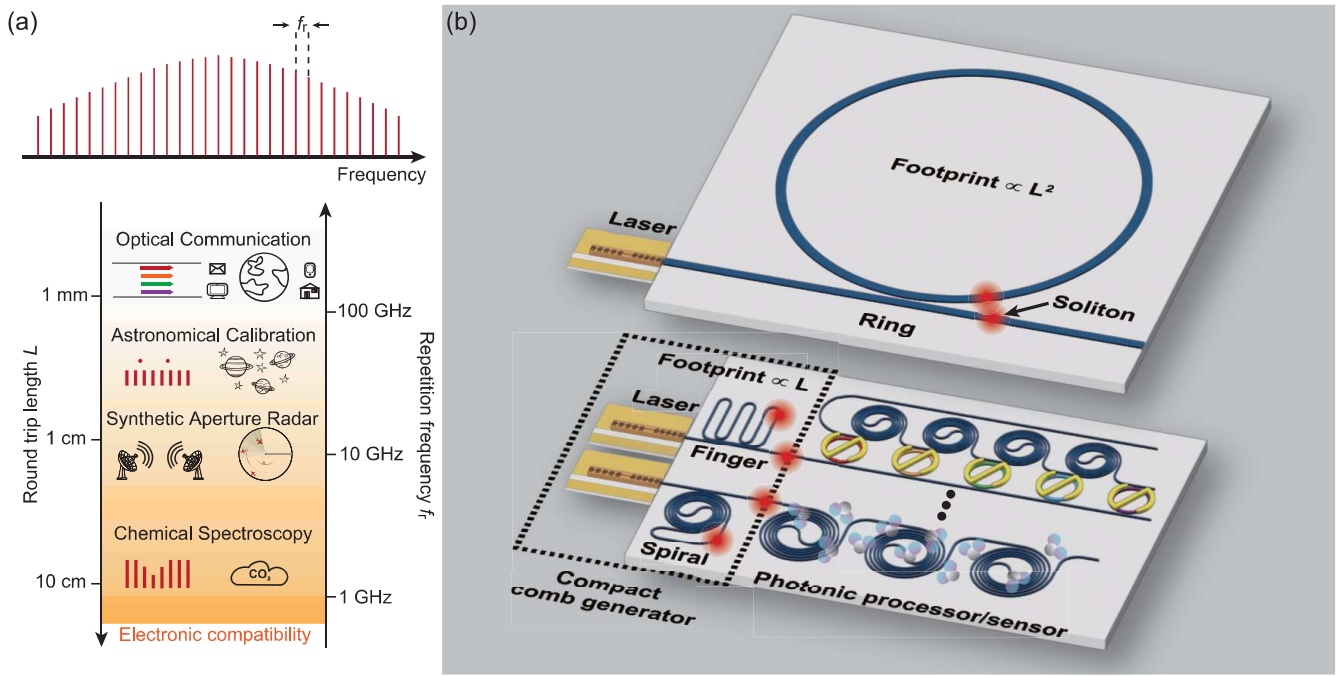

**Fig. 1.** Conceptual schematic depicting applications of soliton microcombs and their integration into compact photonic systems. a) Schematic of an optical frequency comb with evenly spaced spectral lines defined by the repetition rate ($f_r$). Combs with different $f_r$ values are engineered to address diverse application requirements. b) Comparison of footprints for various microresonator geometries. Soliton microcombs are generated by directly coupling self-injection-locked integrated lasers to microresonators. Compact comb generators utilizing finger-shaped and spiral-shaped microresonators create spaces for additional photonic integrated circuits on the same chip.

### Device fabrication

Figure 2a depicts the key fabrication steps, with complete procedures and parameters detailed in **Experimental Section**. First, stress-release patterns are etched into the 4 μm-thick wet thermal oxide to reduce internal stress in the $Si_3N_4$ film, subdividing the wafer into regions of 5 mm × 1 cm (Figure 2b)[36–41]. Next, $Si_3N_4$ is deposited by low-pressure chemical vapor deposition (LPCVD) to a final thickness of 800 nm, using stepwise and angled processes to balance film stress[41–46]. Electron beam lithography (EBL) defines the resist patterns, which are transferred into the $Si_3N_4$ layer by inductively coupled plasma reactive ion etching (ICP-RIE). To minimize hydrogen incorporation[47,48], the wafer is annealed at 1200 °C in $N_2$ for 10 hours[41,45,46,49,50], reducing the $Si_3N_4$ thickness by approximately 3%. Subsequently, a 1 μm $SiO_2$ cladding is deposited via LPCVD with tetraethyl orthosilicate (TEOS), followed by a second annealing step at 1200 °C to densify the $SiO_2$[15,51]. An additional 2 μm-thick $SiO_2$ layer is then deposited by plasma-enhanced chemical vapor deposition (PECVD). Finally, platinum heaters are formed through a lift-off process[17,52].

The fully processed wafer is diced into $Si_3N_4$ photonic chips measuring 5 mm × 5 mm (Figure 2c), enabling up to 48 chips per 4-inch wafer. Scanning electron microscope (SEM) cross-sections (Figure 2d) confirm the waveguide sidewall angle of 86°, and high-aspect-ratio tapered facet couplers[53] with the width of 240 nm can also be produced.

### Microresonator characterization

We present three types of microresonators, each confined within a 1 mm × 1 mm footprint: microring resonators with FSRs of 100 GHz (Figure 3a), finger-shaped resonators[34] with FSRs of 25 GHz (Figure 3b), and spiral-shaped resonators[34,35,54] with FSRs of 10 GHz (Figure 3c). Microring resonators exhibit lower bending and transition losses and smaller mode crossing due to their constant bending radius. However, for identical FSRs, microring resonators possess larger footprints, rendering them unsuitable for high-density integration in photonic integrated circuits. In contrast, finger-shaped and spiral-shaped resonators achieve higher integration densities as they comprise straight waveguides and Euler bends[55–57], featuring a minimal bending radius of 70 μm. Moreover, spiral-shaped resonators offer even greater integration densities than finger-shaped resonators and can achieve FSRs below 25 GHz within the footprint of 1 mm × 1 mm. Typical $TE_{00}$ transmission spectra for each resonator type at waveguide widths of 2 μm and 4 μm are presented in Figure 3d, e and f, respectively, and are labeled with the loaded linewidths ($\kappa/2\pi$). By fitting these spectra and accounting for resonance doublet effects[58] (details shown in **Experimental Section**), we obtain the intrinsic linewidths ($\kappa_0/2\pi$). The histograms of these intrinsic linewidths are shown in Figure 3g, h and i. For the microring resonators, the most probable $Q_0$ are $1 \times 10^7$ (corresponding to 3.6 dB m$^{-1}$) for the waveguide width of 2 μm and $1.76 \times 10^7$ (2.1 dB m$^{-1}$) for the width of 4 μm. Finger-shaped resonators exhibit similar values of $Q_0$. Spiral-shaped resonators exhibit slightly lower values of $Q_0$, with the most probable values being $9.3 \times 10^6$ (4.0 dB m$^{-1}$) for the 2 μm width and $1.49 \times 10^7$ (2.5 dB m$^{-1}$) for the 4 μm width, respectively. These results indicate negligible bending and transition losses in finger-shaped resonators, whereas losses associated with S-bends in spiral resonators are non-negligible[55]. Consequently, for FSRs exceeding 25 GHz, finger-shaped resonators are preferred, while spiral-shaped resonators are the better choice for FSRs below 25 GHz. This selection criterion ensures minimal loss and better performance across different FSR regimes, facilitating the application of these microresonators in various photonic systems.

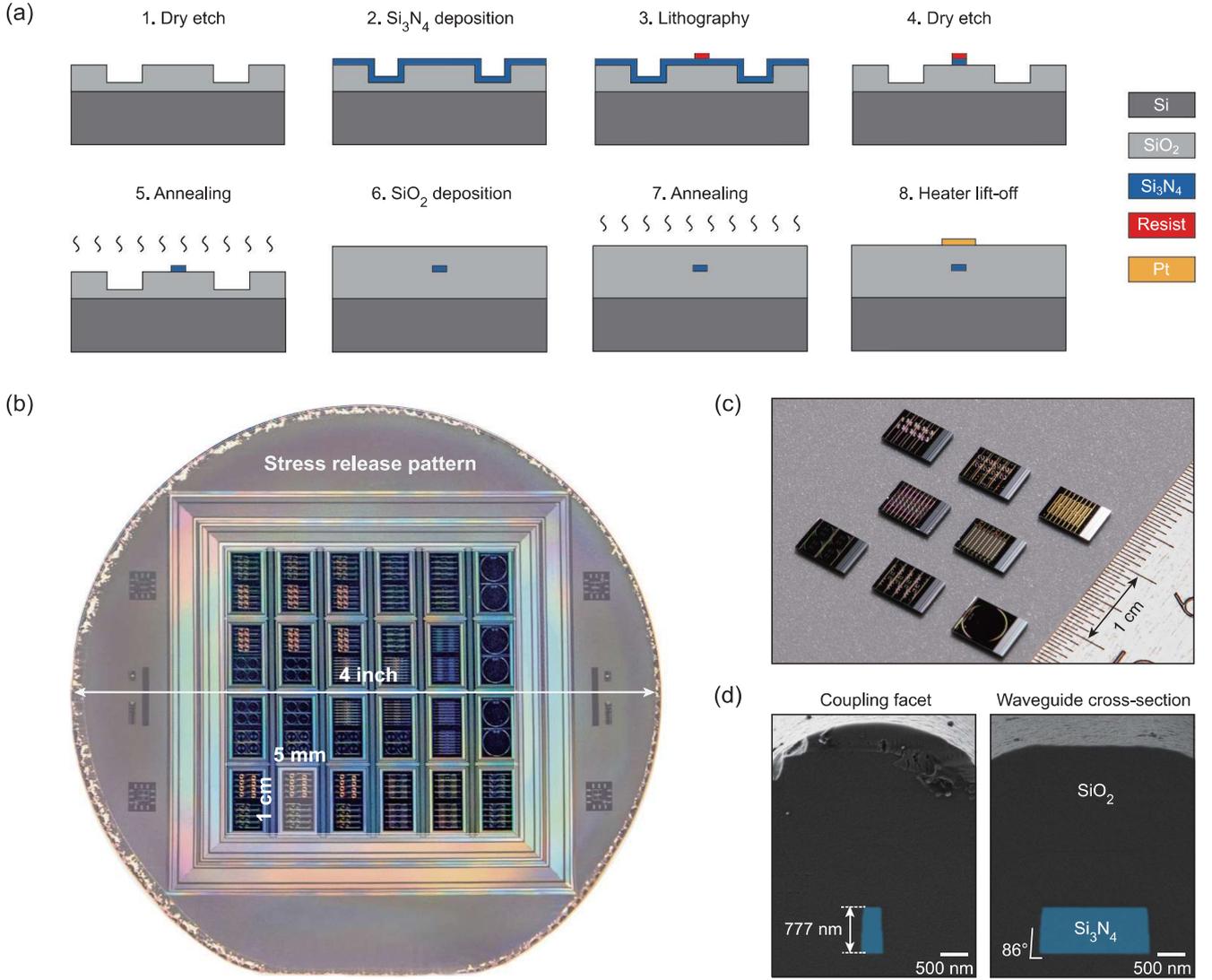

**Fig. 2.** Wafer-scale fabrication of $Si_3N_4$ microresonators. a) Fabrication process flow. b) Photograph of the processed 4-inch wafer. c) Photograph of diced photonic chips. d) SEM cross-sectional images: left, coupling facet (240 nm width); right, waveguide (1.8 μm width).

Anomalous group velocity dispersion (GVD), which is essential for soliton microcomb generation[2,4,5], is quantified by the integrated dispersion $D_{\text{int}}(\mu) = \omega_\mu - \omega_0 - D_1\mu = \sum_{n>2} \frac{D_n\mu^n}{n!}$, where $\omega_\mu/2\pi$ and $\omega_0/2\pi$ denote the $\mu$th and central resonance frequencies, respectively; $\mu$ is the mode index relative to the central mode, and $D_1/2\pi$ represents the resonator FSR. Anomalous GVD corresponds to the positive $D_2$. We measure $D_{\text{int}}$ using a tunable laser calibrated via a Mach–Zehnder interferometer[7]. The results are summarized in Figure 3j, k and l. All resonators exhibit anomalous GVD, with those having a waveguide width of 2 μm demonstrating higher $D_2$ values compared to those with a width of 4 μm. Further increasing the waveguide width may eventually result in normal GVD.

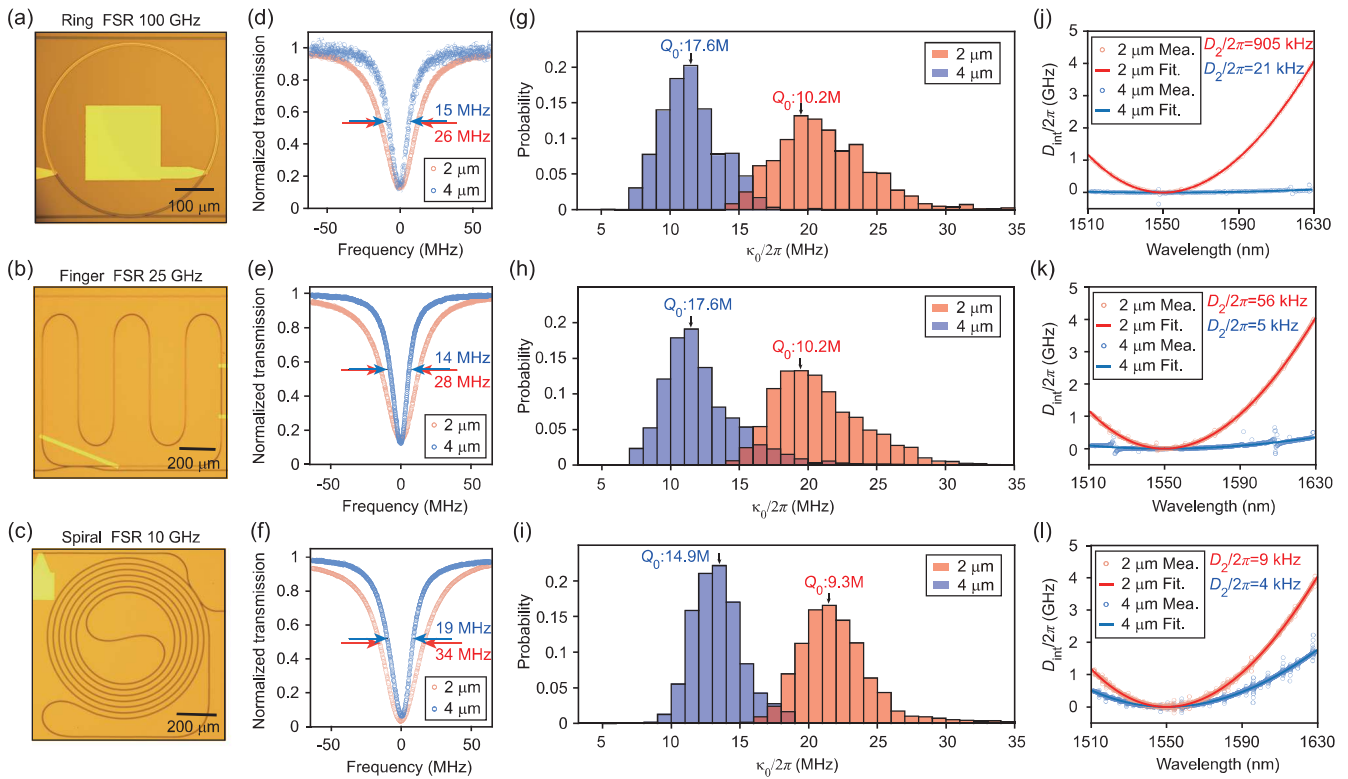

**Fig. 3.** Characterization of microring, finger-shaped, and spiral-shaped resonators. a,b,c Optical microscope images of a 100 GHz microring resonator (a), a 25 GHz finger-shaped resonator (b), and a 10 GHz spiral-shaped resonator (c), respectively. d,e,f Typical TE$_{00}$ resonance profiles for 100 GHz microring resonators (d), 25 GHz finger-shaped resonators (e), and 10 GHz spiral-shaped resonators (f), respectively. Measurements were conducted on microresonators with waveguide widths of 2 μm (red) and 4 μm (blue), with the corresponding values of $\kappa/2\pi$ indicated. g,h,i Distribution histograms of $\kappa_0/2\pi$ for 100 GHz microring resonators (g), 25 GHz finger-shaped resonators (h), and 10 GHz spiral-shaped resonators (i), respectively. Microresonators with waveguide widths of 2 μm (red) and 4 μm (blue) are shown, with the most probable $Q_0$ indicated. j,k,l Measured integrated dispersion for 100 GHz microring resonators (j), 25 GHz finger-shaped resonators (k), and 10 GHz spiral-shaped resonators (l), respectively. The dispersion of microresonators with waveguide widths of 2 μm (red) and 4 μm (blue) is fitted using $D_{\rm int}(\mu) = D_2\mu^2/2$, with the fitted values of $D_2/2\pi$ indicated.

We investigate the intrinsic losses of these microresonators by decomposing them into material absorption ($\kappa_{\rm abs}$) and scattering ($\kappa_{\rm scatter}$) components[14,52,57,59]. Since absorbed power induces heating of the device, leading to a thermally induced redshift of the resonance[59,60], as shown in Figure 4a. We can extract $\kappa_{\rm abs}$ by fitting the shifted transmission spectra (**Experimental Section**). Subsequently, $\kappa_{\rm scatter}$ is obtained by subtracting $\kappa_{\rm abs}$ from $\kappa_0$. As illustrated in Figure 4b, for the TE$_{00}$ mode of a 5 μm-wide microresonator, $\kappa_{\rm abs}/2\pi$ exceeds 10 MHz within the 1520-1540 nm wavelength range, attributed to Si–H and N–H bonds, while a broad 6 MHz background persists at other wavelengths. These absorption linewidths exceed those reported in Reference[57] where $\kappa_{\rm abs}/2\pi \approx 1$ MHz. We attribute this discrepancy primarily to ultraviolet exposure during the PECVD process, which could be mitigated by incorporating an additional annealing step[41]. In comparison, $\kappa_{\rm scatter}/2\pi$ remains constant at 3 MHz across this wavelength range.

Figure 4c presents $\kappa_{\rm scatter}$ for microring resonators with various widths. For waveguide widths below 3 μm, $\kappa_{\rm scatter}$ decreases rapidly with increasing width due to reduced mode overlap with the sidewalls. The sidewall roughness predominantly originates from lithography rather than etching, as indicated by the SEM image (inset of Figure 4c), which shows that the sidewall stripes correspond to the resist pattern. For widths exceeding 3 μm, further widening results in only marginal reductions in scattering, suggesting that top and bottom surface roughness may become the dominant sources of the scattering loss. These losses can be mitigated by applying chemical–mechanical polishing to the Si$_3$N$_4$ film[14,52,57,61,62].

## Turnkey soliton microcombs

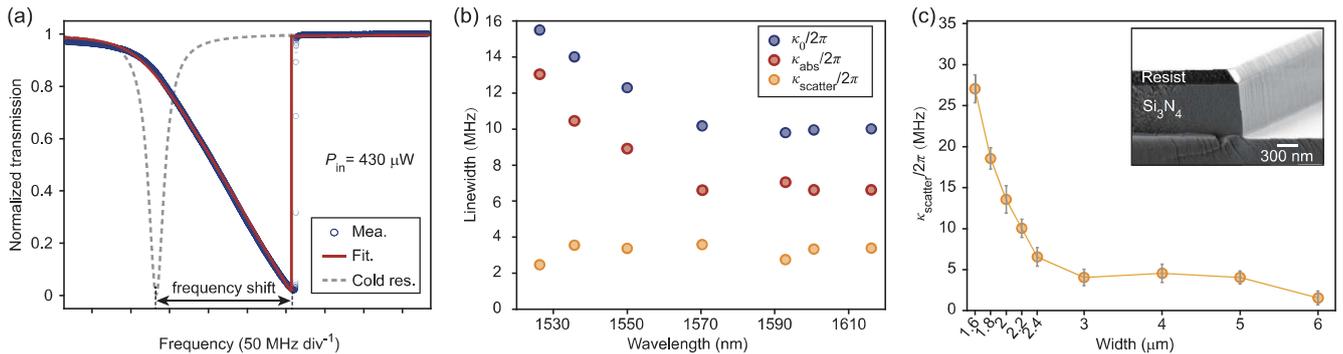

**Fig. 4.** Characterization of absorption and scattering losses in microresonators. a) Transmission spectra of a resonance for 430 μW optical power (blue circles) and much lower power (gray dashed line) in the bus waveguide. The red line is a fit using the model in **Experimental Section**. b) Wavelength dependence of the intrinsic loss rate ($\kappa_0$), absorption loss rate ($\kappa_{abs}$), and scattering loss rate ($\kappa_{scatter}$) for a microresonator with a waveguide width of 5 μm. c) Scattering loss rate as a function of microresonator waveguide width. Gray error bars indicate the standard deviation. Inset: SEM image showing the sidewall of the etched Si₃N₄ waveguide and the resist.

We demonstrate turnkey soliton microcombs by directly coupling commercial DFB lasers to Si₃N₄ chips[19] (Figure 5a). In the absence of an isolator within the setup, backscattering within the microresonator is reintroduced into the laser, thereby altering its tuning curve. This phenomenon, known as self-injection locking[6,18,19,21], can lead to a reduced laser linewidth and facilitate the generation of soliton microcombs when the feedback phase is appropriately adjusted. In our experimental configuration, the feedback phase is controlled using a piezoelectric stage. By injecting 30 mW of optical power into the bus waveguide and optimizing the feedback phase, we observe the formation of microcombs as the laser frequency is tuned into resonance (Figure 5b). Notably, a low-noise step emerges in the comb power traces, corresponding to the formation of a single soliton within the microresonator. The soliton state persists over a laser current range of up to 7 mA, and by maintaining the laser current within this range, soliton microcombs can be reliably accessed. Consequently, we observe single-soliton microcombs with $f_r$ of 100 GHz, 25 GHz, and 10 GHz (Figure 5c), each spanning over 100 nm in optical spectra.

To assess the coherence of the soliton microcombs, we measure their electrical beatnotes corresponding to $f_r$ using an electrical spectral analyzer. The residual pump light of the soliton microcombs is removed using a notch filter. The soliton microcomb is then amplified to 2 mW before sending into a high-speed photodetector. The electrical beatnotes of the 25 GHz and 10 GHz soliton microcombs manifest as stable, single-peaked signals, a signature of mode-locking (Figure 5d). Further characterization of the phase noise using a phase noise analyzer is presented in Figure 5e. Notably, the 10 GHz soliton's $f_r$ exhibits a phase noise of -130 dBc Hz⁻¹ at a 100 kHz offset. At higher offset frequencies, the phase noise is primarily limited by the amplified spontaneous emission noise of the optical amplifier, which can be mitigated by increasing the output power of the soliton microcomb itself.

The soliton microcombs are encapsulated within a standard butterfly package, ensuring a fixed feedback phase and reliable operation. To emulate the turn-on process of the soliton microcomb, we modulate the laser current with a square wave (Figure 6a). The results demonstrate that each time the laser current is adjusted to a predetermined value, a single soliton microcomb is deterministically generated. We then evaluate the long-term operational stability. With temperature control, the power of the 10 GHz soliton microcomb exhibits fluctuations of less than 5 % over 5000 s (Figure 6b). Additionally, the electrical beatnote drifts within 2 kHz for durations exceeding one hour (Figure 6c). This corresponds to a fractional Allan deviation below $10^{-8}$ for averaging times up to 100 s (Figure 6d), although it may increase at longer timescales due to other sources of drift. The butterfly package can be further integrated into a module measuring 10 cm × 7 cm × 4 cm, incorporating the entire drive current source and temperature control units onto an electronic circuit (Figure 6e). The overall size is comparable to that of a computer mouse, which is suitable for operation outside laboratory environments.

## Conclusion and outlook

In conclusion, we demonstrate wafer-scale fabrication of high-$Q$ Si₃N₄ microresonators, including compact finger-shaped and spiral-shaped designs with intrinsic quality factors ($Q_0$) exceeding $10^7$. The resulting soliton microcombs reliably cover microwave repetition rates ranging from 10 GHz to 100 GHz, achieving phase noise levels as low as -130 dBc Hz⁻¹ at a 100 kHz offset, which ranks among the lowest reported for fully integrated microcombs[21,63]. Crucially, these devices are fully packaged and exhibit high technological readiness, immediately unlocking numerous applications in practical scenarios. For instance, in radar systems, they can en-hance target resolution and detection range by providing low-phase-noise local oscillators[64]; in sensing, they can actively track multiple chemical species with broadband spectral access[65,66]; and they can serve as multiplexed light sources in communications by offering massive data-transmission bandwidth with high-spectral efficiency[28].

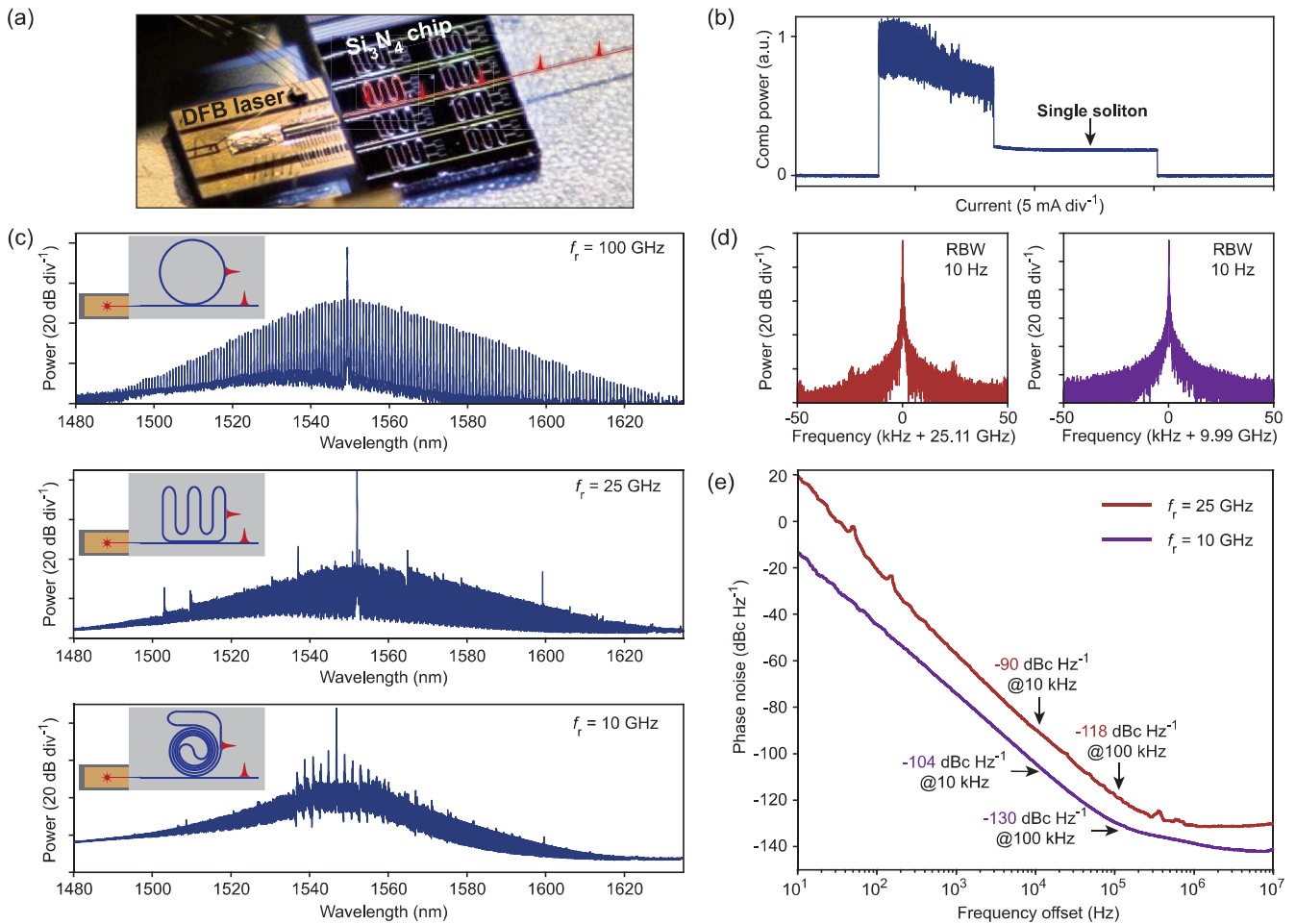

**Fig. 5.** Soliton microcombs pumped by DFB lasers. a) Experimental setup showing a DFB laser directly coupled to a $Si_3N_4$ chip. b) Measured comb power as a function of laser current scanning. c) Optical spectra of single-soliton microcombs at $f_r$ of 100 GHz, 25 GHz, and 10 GHz, from top to bottom. d) Electrical beatnotes of 25 GHz (left panel) and 10 GHz (right panel) soliton microcombs. e) Single-sideband phase noise of the electrical beatnotes of 25 GHz (red) and 10 GHz (purple) soliton microcombs.

At present, our modules house only the soliton microcomb itself. Looking ahead, the compact geometries of these high-$Q$ microresonators facilitate high-density integration with additional on-chip elements, such as lasers[20,67,68], nonlinear waveguides[2,69,70], amplifiers[71,72], electro-optic modulators[73–76], spectrometers[77] and photodetectors[78–80]. Such robust photonic integration is expected to reduce insertion losses, suppress crosstalk, and enhance energy efficiency. Moreover, large-volume manufacturing can further reduce costs, accelerating the transition of microcomb technologies from academic demonstrations to widespread industrial adoption. This progress holds promise for a new era of ultra-compact, cost-effective, and high-performance coherent photonic systems.

## Experimental Section
### Fabrication Details

Stress-release patterns are defined using S1818 photoresist and a contact ultraviolet lithography system (SUSS MA/BA6). These patterns are subsequently etched into the oxide layer to a depth of 3 µm using ICP-RIE with a gas mixture of $CHF_3$ and $O_2$. The stoichiometric $Si_3N_4$ film is deposited in an LPCVD furnace, utilizing $SiCl_2H_2$ and $NH_3$ precursors at a reaction temperature of 770 °C, achieving a deposition rate of approximately $3\,nm\,min^{-1}$. To obtain the target thickness of 800 nm, an initial 400 nm of $Si_3N_4$ is deposited, followed by a 45-degree wafer rotation before depositing the remaining 400 nm[41–46].

EBL is performed using a Raith 5200 system to ex-pose ma-N 2405 resist with a thickness of approximately 720 nm. The EBL process employs a beam step size of 5 nm and a beam current of 5 nA to deliver a dose of approximately $400\,\mu\mathrm{C\,cm^{-2}}$. To minimize surface roughness, a shot arrangement technique[14] is utilized. Additionally, the microresonators are positioned within a single writing field (1 mm × 1 mm) whenever possible to mitigate stitching effects[81,82].

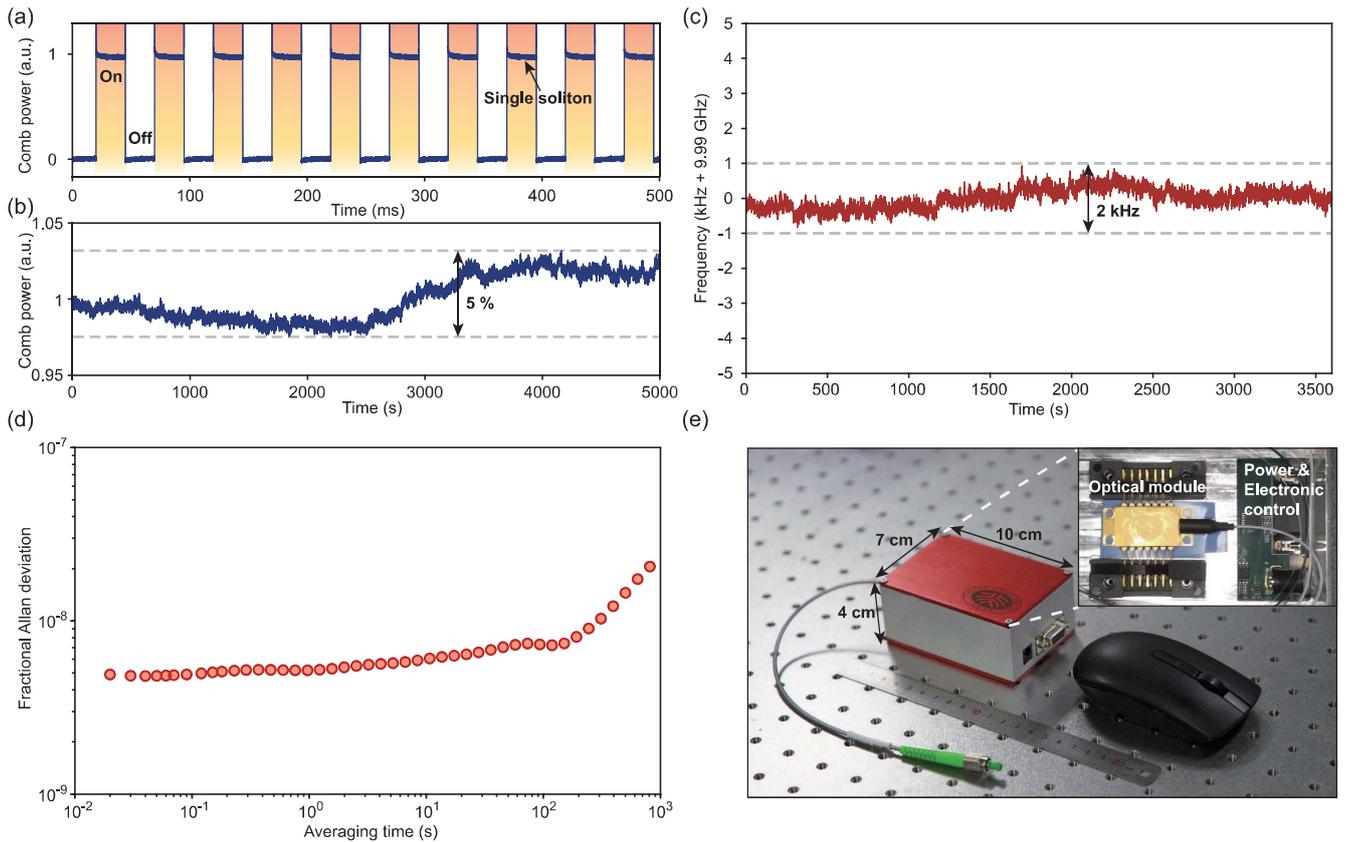

**Fig. 6.** Reliability tests of packaged turnkey soliton microcombs. a) Repeatability of turnkey initiation. Ten consecutive turn-on tests of the soliton microcomb, with the "on" states highlighted by the shadings. b) Measured comb power of a 10 GHz soliton microcomb as a function of time. c) Measured $f_r$ of a 10 GHz soliton microcomb as a function of time. d) Fractional Allan deviation of $f_r$ presented in (c). e) Images of a turnkey soliton microcomb module. Insets: Detailed photographs of the internal optical modules and electrical control circuits.

The $\mathrm{Si_3N_4}$ film is dry etched by ICP-RIE using the same precursors as those used for $\mathrm{SiO_2}$ etching. After annealing and deposition of the $\mathrm{SiO_2}$ cladding, S1813 photoresist is applied and patterned using the contact ultraviolet lithography system (SUSS MA/BA6) to define the metal heater patterns. Subsequently, an electron beam evaporation system (Peva-600E) deposits 10 nm of titanium and 100 nm of platinum, followed by a lift-off process to pattern the heaters[17,52].

**Fitting of Transmission Spectra**
When backscattering occurs within the resonator, an additional backward-propagating field $b$ must be introduced alongside the forward-propagating field $a$. Assuming both fields experience the same intrinsic decay rate

($\kappa_0$) and coupling rate ($\kappa_{\mathrm{ext}}$), their temporal evolutions are governed by:

$$\frac{\mathrm{d}a(t)}{\mathrm{d}t} = -\left[ i(\omega_0 - \omega_{\mathrm{p}}) + \frac{\kappa}{2} \right] a(t) + \sqrt{\kappa_{\mathrm{ext}}}\, s_{\mathrm{in}}(t) + i\,\kappa_{\mathrm{c}}\, b(t),$$

$$\frac{\mathrm{d}b(t)}{\mathrm{d}t} = -\left[ i(\omega_0 - \omega_{\mathrm{p}}) + \frac{\kappa}{2} \right] b(t) + i\,\kappa_{\mathrm{c}}\, a(t),$$
(1)

where $\kappa = \kappa_0 + \kappa_{\mathrm{ext}}$ is the total decay rate, $\omega_0$ and $\omega_{\mathrm{p}}$ denote the resonant frequency and pump frequency, respectively, and $\kappa_{\mathrm{c}}$ is the coupling rate between the forward and backward fields. By neglecting dissipative coupling, the steady-state transmission of the resonator is found to be

$$\mathcal{T}_{\mathrm{Split}} = 1 - \frac{4\,\kappa_{\mathrm{ext}}\left[ 4\,\kappa_{\mathrm{c}}^2\,\kappa + \kappa_0\left(\kappa^2 + 4\,\omega^2\right) \right]}{\left[ \kappa^2 + 4\left(\kappa_{\mathrm{c}} - \Delta\omega\right)^2 \right]\left[ \kappa^2 + 4\left(\kappa_{\mathrm{c}} + \Delta\omega\right)^2 \right]},$$
(2)

where $\Delta\omega = \omega_0 - \omega_{\mathrm{p}}$ represents the detuning between the resonance and the pump frequencies. This expression reveals that a single Lorentzian resonance dip in the transmission spectrum splits into two dips due to the coupling of the forward and backward-propagating fields.

**Characterization of Material Absorption**

When optical power is launched into the resonator, the combined influence of Kerr nonlinearity and the photothermal effect induces a shift in the resonance frequency[59]:

$$\delta\omega_0 = -(\alpha + g)\rho, \tag{3}$$

where $\alpha$ and $g$ denote the photothermal and Kerr coefficients, respectively, and $\rho$ is the intracavity energy density, which are defined as:

$$\alpha = \overline{\kappa_{\mathrm{abs}}} \, \frac{\overline{\delta T}}{P_{\mathrm{abs}}} \left(-\frac{\delta\omega_0}{\delta T}\right) V_{\mathrm{eff}},$$
$$g = \frac{\omega c \overline{n_2}}{\overline{n_0 n_g}}, \tag{4}$$
$$\rho = \frac{|a|^2}{V_{\mathrm{eff}}},$$

where $\kappa_{\mathrm{abs}}$ is the energy decay rate arising from material absorption, $\delta T$ is the temperature rise in the resonator, $P_{\mathrm{abs}}$ is the absorbed optical power, $V_{\mathrm{eff}}$ is the effective mode volume of the pumped resonance, $n_2$ is the material Kerr refractive index, $n_0$ is the material refractive index, and $n_g$ is the material chromatic group refractive index.

Here, the overline denotes an average weighted by the optical field distribution[59], whose weights are given by mode profiles simulated by finite element analysis.

By incorporating these relations into the Lorentzian transmission of the resonator, the resonance shift becomes:

$$\mathcal{T}_{\mathrm{Shifted}} = 1 - \frac{\kappa_{\mathrm{ext}} \, \kappa_0}{\left[\Delta\omega - C_{\mathrm{shift}}\left(1 - \mathrm{T}_{\mathrm{Shifted}}\right)\right]^2 + \left(\frac{\kappa}{2}\right)^2}, \tag{5}$$

where $\mathcal{T}_{\mathrm{Shifted}}$ implicitly depends on $\Delta\omega$. Fitting this expression allows one to extract the resonance shift coefficient:

$$C_{\mathrm{shift}} = \frac{1}{\kappa_0}(\alpha + g) P_{\mathrm{in}}. \tag{6}$$

Finite element simulations show that $\overline{\delta T}/P_{\mathrm{abs}} = 180\,\mathrm{K}\,\mathrm{W}^{-1}$ and $V_{\mathrm{eff}} = 3 \times 10^{-15}\,\mathrm{m}^3$. Taking $n_0$ and $n_g$ for $\mathrm{SiO}_2$ as $n_0 = 1.44, n_g = 1.46$, for $\mathrm{Si}_3\mathrm{N}_4$ as $n_0 = 2.00, n_g = 2.04$[83,84], the weighted averages of $\overline{n_0 n_g}$ and $\overline{n_2}$ are determined to be $\overline{n_0 n_g} = 3.91, \overline{n_2} = 2.18 \times 10^{-19}$ for the 5 µm-wide microring. The resonance frequency shift per unit temperature change, $\delta\omega_0/\delta T$, is experimentally determined to be approximately $2\pi \times 3.7\,\mathrm{GHz}\,\mathrm{K}^{-1}$. Substituting these parameters into the above equations enables the extraction of the absorption rate $\kappa_{\mathrm{abs}}$.

## Acknowledgments

This work was supported by National Key R&D Plan of China (Grant No. 2023YFB2806702), Beijing Natural Science Foundation (Z210004, Z240007), and National Natural Science Foundation of China (12293051,92150108). The authors thank Junqiu Liu, Shizhuoluo Wang, Wenjing Liu, Zhendong Zhu for their insightful discussions, Jincheng Li, Zhigang Hu, Hao Yang, Ruokai Zheng, Xiaoxuan Peng for assistance in fabrication, Jiadong Wang, Jun Mao, Du Qian for assistance on sample photography, and Binbin Nie, Xinrui Luo, Yiheng Li, Du Qian, Qixuan Zhou, Hanfei Hou, Shenyu Xiao for assistance in measurements. The fabrication in this work was supported by the Micro/nano Fabrication Laboratory of Synergetic Extreme Condition User Facility (SECUF), Songshan Lake Materials Laboratory, and the Advanced Photonics Integrated Center of Peking University.

## Competing interests

The authors declare no competing interests.

## Data availability

The data that support the findings of this study are available from the corresponding author upon reasonable request.